\documentclass[aps,twocolumn,showpacs,preprintnumbers,floats]{revtex4}\def\@cite#1#2{\textsuperscript{[{#1\if@tempswa , #2\fi}]}}
\usepackage{mathrsfs}
\usepackage{longtable,lscape}
\usepackage{txfonts}
\usepackage{amssymb}
\usepackage{indentfirst}
\usepackage{graphicx,booktabs}
\usepackage{float}
\usepackage{setspace}

\usepackage{multirow}
\usepackage{color}
\usepackage{epsfig}

\begin{document}


\title{A new decay mode of higher charmonium}

\author{Li-Ye Xiao$^{1,2}$~\footnote {E-mail: lyxiao@pku.edu.cn}, Xin-Zhen Weng$^{1}$, Qi-Fang L\"u$^{3,4}$~\footnote {E-mail: lvqifang@hunnu.edu.cn}, Xian-Hui Zhong$^{3,4}$~\footnote {E-mail: zhongxh@hunnu.edu.cn}, and Shi-Lin Zhu$^{1,2,5}$~\footnote {E-mail: zhusl@pku.edu.cn}}

\affiliation{ 1) School of Physics and State Key Laboratory of
Nuclear Physics and Technology, Peking University, Beijing 100871,
China } \affiliation{ 2)  Center of High Energy Physics, Peking
University, Beijing 100871, China} \affiliation{ 3) Department of
Physics, Hunan Normal University, and Key Laboratory of
Low-Dimensional Quantum Structures and Quantum Control of Ministry
of Education, Changsha 410081, China } \affiliation{ 4) Synergetic
Innovation Center for Quantum Effects and Applications (SICQEA),
Hunan Normal University, Changsha 410081, China} \affiliation{ 5)
Collaborative Innovation Center of Quantum Matter, Beijing 100871,
China}


\begin{abstract}
We calculate the $\Lambda_c\bar{\Lambda}_c$ partial decay width of
the excited vector charmonium states around 4.6 GeV with the quark
pair creation model. We find that the partial decay width of the
$\Lambda_c\bar{\Lambda}_c$ mode can reach up to several MeV for
$\psi(4S,~5S,~6S)$. In contrast, the partial $\Lambda_c\bar{\Lambda}_c$
decay width of the states $\psi(3D,~4D,~5D)$ is less than one MeV.
If the enhancement $Y(4630)$ reported by the Belle Collaboration in
$\Lambda_c\bar{\Lambda}_c$ invariant-mass distribution is the same
structure as $Y(4660)$, the $Y(4660)$ resonance is most likely to be a
$S$-wave charmonium state.


\end{abstract}

\pacs{}

\maketitle

\section{Introduction}
Since the Belle Collaboration reported the first member of the
family of the charmonium-like states, $X(3872)$, in
2003~\cite{Choi:2003ue}, a series of charmonium-like
states~\cite{Patrignani:2016xqp}, called collectively $XYZ$ states,
have been observed by several major experimental
collaborations such as Babar, BESIII, LHCb, CLEO-c and so on. To
date, dozens of charmonium-like states~\cite{Patrignani:2016xqp}
have been discovered. The charmonium systems may provide unique
clues to the nonperturbative behavior of QCD in the low energy
regime and have attracted a great deal of attention from the hadron
physics community; see Ref.~\cite{Chen:2016qju} for a review and
references.

The $Y(4660)$ resonance, as the most massive state among the charmonium-like
states at present, was first reported by the Belle
Collaboration~\cite{Wang:2007ea} in the process $e^+e^-\rightarrow
\gamma_{\text{ISR}}\pi^+\pi^-\psi(2S)$ associated with the $Y(4360)$ resonance
in 2007. Its mass and width were determined to be $M=(4664\pm16)$
MeV and $\Gamma=(48\pm18)$ MeV, respectively. Later, this state was
confirmed by BaBar collaboration~\cite{Lees:2012pv} in the
$\pi^+\pi^-\psi(2S)$ invariant-mass distribution with new data on
the $e^+e^-\rightarrow \gamma_{\text{ISR}}\pi^+\pi^-\psi(2S)$
progress with the mass $M=(4669\pm24)$ MeV and width
$\Gamma=(104\pm58)$ MeV. Since the $Y(4660)$ resonance was produced
from the $e^+e^-$ annihilation, the quantum number is
$J^{PC}=1^{--}$. Besides, the Belle Collaboration reported an
enhancement, $Y(4630)$, in the cross section of the $e^+e^-\rightarrow
\Lambda^+_c\Lambda^-_c$ in 2008~\cite{Pakhlova:2008vn}, whose mass
and width are consistent within the errors with those of
$Y(4660)$. Hence, these two states may be the same structure
although they were observed in different
processes~\cite{Bugg:2008sk,Cotugno:2009ys,Guo:2010tk}.

Over the past decade, the properties of the charmonium-like state
$Y(4660)$ were extensively explored with various theoretical
methods. In the framework of the screened potential model by Li and
Chao~\cite{Li:2009zu}, $Y(4660)$ was a good candidate of the
$\psi(6S)$ state. However, Ding et al.~\cite{Ding:2007rg}
interpreted $Y(4660)$ as the $\psi(5S)$ state in the flux tube
model, which is consistent with the prediction in
Ref.~\cite{Gui:2018rvv}. Besides the interpretation of the
conventional charmonium states, $Y(4660)$ was also interpreted
as a tetraquark
state~\cite{Maiani:2014aja,Ebert:2008kb,Wang:2018rfw,Lu:2016cwr,Chen:2010ze,Zhang:2010mw,Albuquerque:2008up,
Wang:2013exa,Wang:2016mmg,Sundu:2018toi}, $f_0(980)\psi'$ bound
state~\cite{Guo:2008zg,Guo:2009id,Wang:2009hi,Albuquerque:2011ix},
baryonium~\cite{Cotugno:2009ys,Qiao:2007ce} and
hadro-charmonium state~\cite{Dubynskiy:2008mq} and so on. In
addition, van Beveren \emph{et al}.~\cite{vanBeveren:2010jz} argued that
$Y(4660)$ should not be associated with a resonance pole of the
$c\bar{c}$ propagator by analyzing the published Babar data for the
reaction $e^+e^-\rightarrow D^*\bar{D}^*$ ~\cite{Aubert:2009aq}. For the properties of $Y(4630)$, there are many theoretical interpretations as well~\cite{Liu:2016nbm,Liu:2016sip,Guo:2016iej,Wang:2016fhj,Lee:2011rka,Simonov:2011jc,vanBeveren:2008rt}.

According to the mass and spin-parity, the possible assignments of
$Y(4660)$ as a charmonium state are $\psi(4S)$, $\psi(5S)$,
$\psi(6S)$, $\psi(3D)$, $\psi(4D)$, or $\psi(5D)$, which have been
listed in Table~\ref{MASS}. In the framework of the quark pair
creation model (QPC model), we calculate the decay width of the
$\Lambda_c\bar{\Lambda}_c$ mode for those vector charmonium states
and obtain that (i) if the $Y(4660)$ is a $S$-wave charmonium state,
the partial decay width of the $\Lambda_c\bar{\Lambda}_c$ mode can
reach several MeV. However, if the $Y(4660)$ is a $D$-wave
charmonium state, the partial decay width of the
$\Lambda_c\bar{\Lambda}_c$ mode should be less than one MeV. (ii) If
the enhancement $Y(4630)$ reported by Belle Collaboration in
$\Lambda_c\bar{\Lambda}_c$ invariant-mass distribution is the same
structure as $Y(4660)$, the $Y(4660)$ resonance is most likely to be a
$S$-wave charmonium state.

This paper is organized as follows. In Sec. II we give a brief
introduction of the QPC model. Then, we present our numerical
results and discussions in Sec. III and summarize our results in
Sec. IV.


\begin{table*}[htpb]
\caption{\label{MASS} The possible assignments of the $Y(4660)$ with
the predicted masses (MeV) from various models. }
\begin{tabular}{ccccccccccc}\hline\hline
\text{State}
~~~~~~~~&\text{QM}~\cite{Eichten:1979ms}~~~~~~~~&\text{QM}~\cite{Godfrey:1985xj}~~~~~~~~&\text{QM}~\cite{Segovia:2008ta}~~~~~~~~&\text{SSE}/\text{EA}\cite{Badalian:2008dv}~~~~~~~~&\text{NR}/\text{GI}~\cite{Barnes:2005pb}
~~~~~~~~&\text{SP}~\cite{Li:2009zu} ~~~~~~~~&\text{LP}/\text{SP}~\cite{Deng:2016stx}\\
\hline
$\psi(4^3S_1)$~~~~~~~~&4625~~~~~~~~&4450~~~~~~~~&4389~~~~~~~~&4398/4426~~~~~~~~&4406/4450~~~~~~~~&4273~~~~~~~~&4412/4281 \\
$\psi(5^3S_1)$~~~~~~~~&$\cdot\cdot\cdot$~~~~~~~~&$\cdot\cdot\cdot$~~~~~~~~&4641~~~~~~~~&4642/4672~~~~~~~~&$\cdot\cdot\cdot$~~~~~~~~&4463~~~~~~~~&4711/4472 \\
$\psi(6^3S_1)$~~~~~~~~&$\cdot\cdot\cdot$~~~~~~~~&$\cdot\cdot\cdot$~~~~~~~~&$\cdot\cdot\cdot$~~~~~~~~&4804/4828~~~~~~~~&$\cdot\cdot\cdot$~~~~~~~~&4608~~~~~~~~&$\cdot\cdot\cdot$ \\
$\psi(3^3D_1)$~~~~~~~~&$\cdot\cdot\cdot$~~~~~~~~&4520~~~~~~~~&4426~~~~~~~~&4464/4477~~~~~~~~&$\cdot\cdot\cdot$~~~~~~~~&4317~~~~~~~~&4478/4336 \\
$\psi(4^3D_1)$~~~~~~~~&$\cdot\cdot\cdot$~~~~~~~~&$\cdot\cdot\cdot$~~~~~~~~&4641~~~~~~~~&4690/4707~~~~~~~~&$\cdot\cdot\cdot$~~~~~~~~&$\cdot\cdot\cdot$~~~~~~~~&$\cdot\cdot\cdot$ \\
$\psi(5^3D_1)$~~~~~~~~&$\cdot\cdot\cdot$~~~~~~~~&$\cdot\cdot\cdot$~~~~~~~~&$\cdot\cdot\cdot$~~~~~~~~&4840/4855~~~~~~~~&$\cdot\cdot\cdot$~~~~~~~~&$\cdot\cdot\cdot$~~~~~~~~&$\cdot\cdot\cdot$ \\
\hline\hline
\end{tabular}
\end{table*}

\section{A introduction of the $^3P_0$ model}\label{model}

The QPC model is known as $^3P_0$ model, which was first proposed by
Micu~\cite{Micu:1968mk}, Carlitz and
Kislinger~\cite{Carlitz:1970xb}, and further developed by the Orsay
group~\cite{LeYaouanc:1972vsx,LeYaouanc:1988fx,LeYaouanc:1977fsz}.
This model is widely used to study the OZI-allowed strong decays of
hadrons. In the model, a quark pair $q\bar{q}$ is created from
the vacuum and then regroups with the quarks within the initial
hadron to produce two outgoing hadrons. In particular, the
interaction Hamiltonian for one quark pair creation was assumed
as~\cite{Geiger:1994kr,Ackleh:1996yt,Close:2005se}
\begin{eqnarray}
H_{q\bar{q}}=\gamma\sum_f2m_f\int d^3x\bar{\psi}_f\psi_f,
\end{eqnarray}
where $m_f$ is the constituent quark mass of flavor $f$, $\psi_f$
denotes a Dirac quark field, and $\gamma$ is a dimensionless
parameter describing the $q\bar{q}$ pair-production strength, which
is usually fixed by fitting the well measured decay widths.

\begin{figure}[htpb]
\centering \epsfxsize=5.8 cm \epsfbox{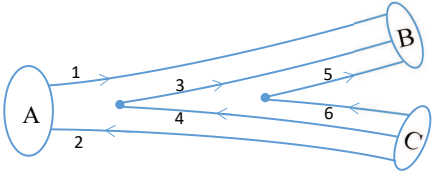} \caption{The
charmonium system decays into a $\Lambda_c\bar{\Lambda}_c$
pair.}\label{qkp}
\end{figure}

In this work, we extend the $^3P_0$ model to study the charmonium
system decaying into a $\Lambda_c\bar{\Lambda}_c$ pair. For this
type of reaction, it is necessary to create two light quark pairs,
which is shown in Fig.~\ref{qkp}. In the framework of the $^3P_0$
model, the helicity amplitude $M^{M_{J_A}M_{J_B}M_{J_B}}$ for the
process of $Y(4660)(A)\rightarrow \Lambda_c(B)+\bar{\Lambda}_c(C)$
reads
\begin{eqnarray}\label{Q1}
\delta^3(\mathbf{p}_A-\mathbf{p}_B-\mathbf{p}_C)M^{M_{J_A}M_{J_B}M_{J_B}}=\frac{\sum_k\langle
BC|H_{q\bar{q}}|k\rangle\langle k|H_{q\bar{q}}|A\rangle}{E_k-E_A}.
\end{eqnarray}
Here, $\mathbf{p}_A(\mathbf{p}_B/\mathbf{p}_C)$ represents the
momentum of the hadron $A(B/C)$. $E_k$ and $E_A$ stand for the
energy of the intermediate state $k$ and initial state $A$,
respectively. To simply the calculations, we take $E_k-E_A$ as a
constant, namely $E_k-E_A\equiv4m_{\mu}$. Under the above
approximation, we can rewrite the Eq.~(\ref{Q1}) as
\begin{eqnarray}\label{Q2}
\delta^3(\mathbf{p}_A-\mathbf{p}_B-\mathbf{p}_C)M^{M_{J_A}M_{J_B}M_{J_B}}=\frac{\langle
BC|H_{q\bar{q}}H_{q\bar{q}}|A\rangle}{4m_{\mu}},
\end{eqnarray}
where $m_{\mu}$ is the reduced mass of the created quark pair.

In the nonrelativistic limit, the transition operator for the two
quark pairs creation under the $^3P_0$ model is given by
\begin{eqnarray}
T&=&\frac{9\gamma^2}{4m_{\mu}} \sum_{m,m'}\langle1m;1-m|00\rangle\langle1m';1-m'|00\rangle\\
\nonumber &&\int
d^3\mathbf{p}_3d^3\mathbf{p}_4d^3\mathbf{p}_5d^3\mathbf{p}_6\delta^3(\mathbf{p}_3+\mathbf{p}_4)\delta^3(\mathbf{p}_5+\mathbf{p}_6)\\
\nonumber &&\times\varphi^{34}_0\omega^{34}_0\chi^{34}_{1,-m}
\mathcal{Y}_1^m(\frac{\mathbf{p}_3-\mathbf{p}_4}{2})a^{\dagger}_{3i}b^{\dagger}_{4j}\\
\nonumber &&\times\varphi^{56}_0\omega^{56}_0\chi^{56}_{1,-m'}
\mathcal{Y}_1^{m'}(\frac{\mathbf{p}_5-\mathbf{p}_6}{2})a^{\dagger}_{5i}b^{\dagger}_{6j},
\end{eqnarray}
where $\mathbf{p}_i$ ($i$=3,~4,~5,~6) corresponds to the
three-vector momentum of the $i$th quark within the two created
quark pairs. $\varphi_0=(u\bar{u}+d\bar{d}+s\bar{s})/\sqrt{3}$ and
$\omega_0=\delta_{ij}$ correspond to the flavor function and color
singlet, respectively. The solid harmonic polynomial
$\mathcal{Y}_1^{m(m')}(\mathbf{p})\equiv|\mathbf{p}|Y^{m(m')}_1(\theta_p,\phi_p)$
stand for the $P$-wave quark pairs, and $\chi_{1,-m(m')}$ are the
spin triplet states for the created quark pairs.
$a^{\dagger}_{i}b^{\dagger}_{j}$ is the creation operator denoting
the quark pairs creation in the vacuum.

Adopting the definition of the mock state~\cite{Hayne:1981zy}, the
meson ($A$) and baryon ($B$) states are defined as, respectively,
\begin{eqnarray}
|A(N_A~^{2S_A+1}L_A{J_AM_{J_A}})(\mathbf{p}_A)\rangle=~~~~~~~~~~~~~~~~~~~~~~~~~~~~~~~~~~~~~~\nonumber \\
\sqrt{2E_A}\varphi^{12}_A\omega^{12}_A\sum_{M_{L_A},M_{S_A}}\langle L_AM_{L_A};S_AM_{S_A}|J_AM_{J_A}\rangle\nonumber \\
\times\int d^3\mathbf{p}_1d^3\mathbf{p}_2\delta^3(\mathbf{p}_1+\mathbf{p}_2-\mathbf{p}_A)\nonumber \\
\times\Psi_{N_AL_AM_{L_A}(\mathbf{p}_1,\mathbf{p}_2)}\chi^{12}_{S_AM_{S_A}}|q_1(\mathbf{p}_1)q_2(\mathbf{p}_2)\rangle,
\end{eqnarray}
\begin{eqnarray}
|B(N_B~^{2S_B+1}L_B{J_BM_{J_B}})(\mathbf{p}_B)\rangle=~~~~~~~~~~~~~~~~~~~~~~~~~~~~~~~~~~~~~~\nonumber \\
\sqrt{2E_B}\varphi^{135}_B\omega^{135}_B\sum_{M_{L_B},M_{S_B}}\langle L_BM_{L_B};S_BM_{S_B}|J_BM_{J_B}\rangle\nonumber \\
\times\int d^3\mathbf{p}_1d^3\mathbf{p}_3d^3\mathbf{p}_5\delta^3(\mathbf{p}_1+\mathbf{p}_3+\mathbf{p}_5-\mathbf{p}_B)\nonumber \\
\times\Psi_{N_BL_BM_{L_B}(\mathbf{p}_1,\mathbf{p}_3,\mathbf{p}_5)}\chi^{135}_{S_BM_{S_B}}
|q_1(\mathbf{p}_1)q_3(\mathbf{p}_3)q_5(\mathbf{p}_5)\rangle.
\end{eqnarray}

\begin{figure}[htpb]
\centering \epsfxsize=3.2 cm \epsfbox{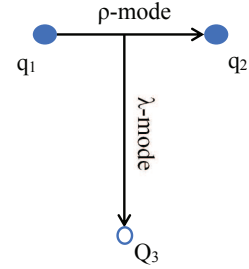} \caption{The
$\rho$- and $\lambda$-mode excitations of the $cqq$ system where
$\rho$ and $\lambda$ are the jacobi coordinates defined as
$\rho=\frac{\mathbf{r}_1-\mathbf{r}_2}{\sqrt{2}}$ and
$\lambda=\frac{\mathbf{r}_1+\mathbf{r}_2-2\mathbf{r}_3}{\sqrt{6}}$,
respectively. $q_1$ and $q_2$ represent the light ($u$, $d$) quark,
and $Q_3$ represents the charm quark.}\label{rlmode}
\end{figure}

The $\mathbf{p}_i~(i=1,2,3,5)$ denotes the momentum of quarks in
hadrons $A$ and $B$. Since the $^3P_0$ model obtains a reasonable
description of the decay properties of many mesons with the simple
harmonic oscillator (SHO) wave functions, and the numerical results
of the decay widths are not strongly sensitive to the details of the
spatial wave
functions~\cite{Kokoski:1985is,Ackleh:1996yt,Geiger:1994kr,Blundell:1995ev},
we adopt the SHO wave functions to describe the space-wave functions of
the baryons in this work. With the simple SHO wave functions, the decay
amplitudes can be calculated analytically. The SHO wave function of
a meson without radial excitations reads
\begin{eqnarray}
\psi^0_{lm}(\mathbf{p})=(-i)^l\Bigg[\frac{2^{l+2}}{\sqrt{\pi}(2l+1)!!}\Bigg]^{\frac{1}{2}}\Bigg(\frac{1}
{\beta}\Bigg)^{l+\frac{3}{2}}
\text{exp}\Bigg(-\frac{\mathbf{p}_R^2}{2\beta^2}\Bigg)\mathcal{Y}_l^m(\mathbf{p}),
\end{eqnarray}
and the ground state space-wave function of a baryon reads
\begin{eqnarray}
\psi_{0,0}=3^{\frac{3}{4}}\Bigg(\frac{1}{\pi\alpha_{\rho}^2}\Bigg)^{\frac{3}{4}}
\Bigg(\frac{1}{\pi\alpha_{\lambda}^2}\Bigg)^{\frac{3}{4}}
\text{exp}\Bigg(-\frac{\mathbf{p}_{\rho}^2}{2\alpha_{\rho}^2}-\frac{\mathbf{p}_{\lambda}^2}{2\alpha_{\lambda}^2}\Bigg).
\end{eqnarray}
Here the $\mathbf{p}_R$ stands for the relative momentum between the
quark and antiquark within the meson. $\mathbf{p}_{\rho}$ and
$\mathbf{p}_{\lambda}$ stand for the momentum corresponding to
$\rho$ and $\lambda$ jacobi coordinates (see Fig.~\ref{rlmode}),
respectively. Thus, we can obtain the helicity amplitude in the
center of mass frame,
\begin{eqnarray}
\mathcal{M}^{M_{J_A}M_{J_B}M_{J_C}}(A\rightarrow B+C)=\frac{9\gamma^2}{4m_{\mu}}\sqrt{8E_AE_BE_C}\nonumber\\
\times\prod_{A,B,C}\langle\chi^{135}_{S_BM_{S_B}}\chi^{246}_{S_CM_{S_C}}|\chi^{12}_{S_AM_{S_A}}\chi^{34}_{1-m}\chi^{56}_{1-m'}\rangle
\nonumber\\
\langle\omega^{135}_B\omega^{246}_C|\omega^{12}_A\omega^{34}_0\omega^{56}_0\rangle
\langle\varphi^{135}_B\varphi^{246}_C|\varphi^{12}_A\varphi^{34}_0\varphi^{56}_0\rangle
I^{M_{L_A},m,m'}_{M_{L_B},M_{L_C}}(\mathbf{p}).
\end{eqnarray}
Here, $I^{M_{L_A},m,m'}_{M_{L_B},M_{L_C}}(\mathbf{p})$ stands the
spatial integral and more detailed calculations are shown in the
Appendix. The $\prod_{A,B,C}$ corresponds to the Clebsch-Gorden
coefficients for the two created quark pairs, initial and final
hadrons, which come from the couplings among the spin, orbital, and
total angular momentum. Its specific expression is
\begin{eqnarray}
\sum\langle L_AM_{L_A};S_AM_{S_A}|J_AM_{J_A}\rangle\langle1m;1-m|00\rangle\langle1m';1-m'|00\rangle\nonumber \\
\times\langle L_BM_{L_B};S_BM_{S_B}|J_BM_{J_B}\rangle\langle
L_CM_{L_C};S_CM_{S_C}|J_CM_{J_C}\rangle.~~~~~
\end{eqnarray}

Finally, the hadronic decay width $\Gamma[A\rightarrow BC]$ reads
\begin{eqnarray}
\Gamma[A\rightarrow
BC]=\pi^2\frac{|\mathbf{p}|}{M_A^2}\frac{1}{2J_A+1}\sum_{M_{J_A},M_{J_B},M_{J_C}}|\mathcal{M}^{M_{J_A}M_{J_B}M_{J_C}}|^2.
\end{eqnarray}
In the equation, the momentum $\mathbf{p}$ of the daughter baryon in
the center of mass frame of the parent baryon $A$ is
\begin{eqnarray}
|\mathbf{p}|=\frac{\sqrt{[M_A^2-(M_B-M_C)^2][M_A^2-(M_B+M_C)^2]}}{2M_A}.
\end{eqnarray}

In the present calculation, we adopt $m_u$=$m_d$=330 MeV and
$m_c$=1628 MeV for the constituent quark masses. The masses of the
baryons $\Lambda_c$ and $\bar{\Lambda}_c$ are taken as
$m_{\Lambda_c}$=$m_{\overline{\Lambda}_c}$=2286.46 MeV, which are
from the Particle Data Group~\cite{Patrignani:2016xqp}. The harmonic
oscillator strength $\beta$ for the excitation between the two charm
quarks in initial charmonium system (see Fig.~\ref{rlmode}) is
adopted as 500 MeV ~\cite{Barnes:2005pb}. The parameter
$\alpha_{\rho}$ of the $\rho$-mode excitation between the two light
quarks in final single-heavy baryons is taken the average value as
$\alpha_{\rho}=400$ MeV, and the other harmonic oscillator parameter
$\alpha_{\lambda}$ is obtained with the following relation~\cite{Wang:2017hej,Xiao:2017udy,Xiao:2017dly}:
\begin{equation}
\alpha_{\lambda}=\Bigg(\frac{3m_Q}{2m_q+m_Q}\Bigg)^{1/4}\alpha_{\rho}.
\end{equation}
In this equation, $m_Q$ stands for the constituent quark mass of
charmed quark, and $m_q$ responds the constituent quark mass of
light quark ($q=u,~d$).

For the strength of the quark pair creation from the vacuum, we
adopt the definition from Ref.~\cite{Segovia:2012cd}, where $\gamma$
is a scale-dependent form,
\begin{equation}\label{Q3}
\gamma(\mu)=\frac{\gamma_0}{\text{log}(\frac{\mu}{\mu_0})}.
\end{equation}
Here, $\mu$ is the reduced mass of the quark-antiquark in the
decaying meson, and $\gamma_0=0.81\pm0.02$ and
$\mu_0=(49.84\pm2.58)$ MeV. According to Eq.~(\ref{Q3}), we get
$\gamma(\mu)=0.29$ with the mass of $m_c=1628$ MeV. So, the strength
of the quark pair creation employed in this work is $\gamma=5.04$
which is $\sqrt{96\pi}$ times of that in Ref.~\cite{Segovia:2012cd}
due to a different
definition~\cite{Godfrey:2015dia,Barnes:2005pb,Ackleh:1996yt}. The
uncertainty of the strength $\gamma$ is around $30\%$ and the
partial decay width is proportional to $\gamma^4$, so the
uncertainty of our theoretical results may be quite large.

\section{Calculations and Results }\label{results}

The quantum number of the $Y(4660)$ resonance is determined to be
$J^P=1^{--}$ from the $e^+e^-$ annihilation. The average values of
mass and width listed in PDG~\cite{Patrignani:2016xqp} are
$M=(4643\pm9)$ MeV and $\Gamma_{\text{total}}=72\pm11$ MeV,
respectively. Around 4660 MeV, there are six vector charmonium
states, which are $\psi(4S)$, $\psi(5S)$, $\psi(6S)$, $\psi(3D)$,
$\psi(4D)$, and $\psi(5D)$. In the following, we will discuss the
decay properties of these states.

\subsection{$S$ wave}

The theoretical mass of $\psi(5^3S_1)$ is about 4.64 GeV (see
Table~\ref{MASS}), which is agreement with the mass of $Y(4660)$ in
PDG~\cite{Patrignani:2016xqp} well. In addition, via evaluating the
open flavor strong decays, some people interpreted $Y(4660)$ as a $\psi(5S)$ state
in the flux tube model~\cite{Ding:2007rg} and QPC model~\cite{Gui:2018rvv}, respectively. As the possibel
assignment of $Y(4660)$, it is crucial to study the decay properties
of the $\psi(5^3S_1)$.

\begin{table}[htpb]
\caption{\label{width} The $\Lambda_c\bar{\Lambda}_c$ partial decay
widths (MeV) of the vector charmonium with a mass of $M=4643$ MeV.
}
\begin{tabular}{ccccccccccc}\hline\hline
\text{State}  ~~~~~~~~~&$\psi(4^3S_1)$~~~~~~~~~&$\psi(5^3S_1)$~~~~~~~~~&$\psi(6^3S_1)$\\
$\Gamma_{\Lambda_c\bar{\Lambda}_c}$~~~~~~~~~&6.57~~~~~~~~~&2.44~~~~~~~~~&0.84 \\
\hline\hline
\text{State}  ~~~~~~~~~&$\psi(3^3D_1)$~~~~~~~~~&$\psi(4^3D_1)$~~~~~~~~~&$\psi(5^3D_1)$\\
$\Gamma_{\Lambda_c\bar{\Lambda}_c}$~~~~~~~~~&0.33~~~~~~~~~&0.19~~~~~~~~~&0.09 \\
\hline\hline
\end{tabular}
\end{table}

According to our calculations, we get
\begin{eqnarray}
\Gamma[\psi(5^3S_1)\rightarrow\Lambda_c\bar{\Lambda}_c]\sim2.44~\text{MeV}
\end{eqnarray}
with a mass of $M=4643$ MeV (see Table~\ref{width}). Combing the measured width of $Y(4660)$, we further predict the
branching ratio
\begin{eqnarray}
\mathcal{B}[\psi(5^3S_1)\rightarrow\Lambda_c\bar{\Lambda}_c]\sim3\%.
\end{eqnarray}
The sizeable branching ratio indicates that this state has a good
potential to be observed in the $\Lambda_c\bar{\Lambda}_c$ decay
channel.

\begin{figure*}[htpb]
\centering \epsfxsize=15.5 cm \epsfbox{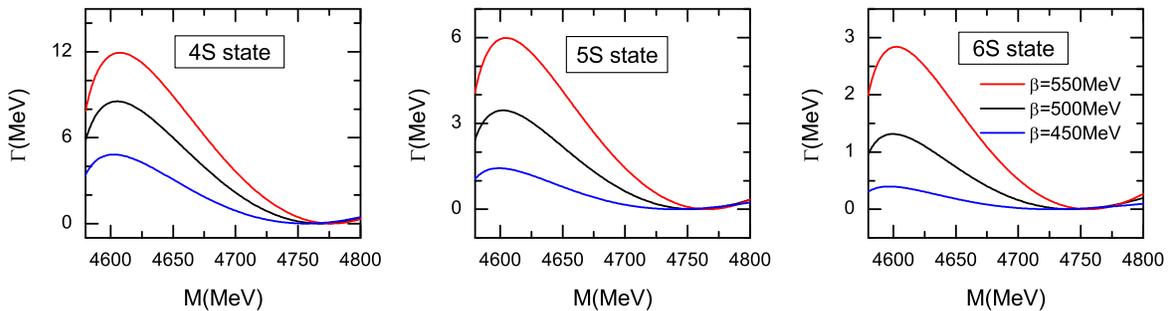} \caption{The
variation of the $\Lambda_c\overline{\Lambda}_c$ decay width with
the mass of the S-wave vector charmonium. The blue, black, and red
lines correspond to the predictions with different values of the
harmonic oscillator strength $\beta=450$, 500, and 550 MeV,
respectively. }\label{SWAVE}
\end{figure*}

Meanwhile, considering the uncertainties of the predicted mass, we
plot the variation of the partial decay width of the
$\Lambda_c\bar{\Lambda}_c$ mode as a function of the mass of the
state $\psi(5^3S_1)$ in Fig.~\ref{SWAVE}. The decay width of the
$\Lambda_c\bar{\Lambda}_c$ mode increases with the mass increasing
in the range of $(4580-4602)$ MeV, and the width can reach up to
$\Gamma\sim3.4$ MeV. However, when the mass increases in the range
of $(4603-4800)$ MeV, the partial decay width decreases.

Besides $\psi(5^3S_1)$, we also investigate the decay properties of
the states $\psi(4^3S_1)$ and $\psi(6^3S_1)$. The predicted masses
of these two states are listed in Table~\ref{MASS}. From the table
it is seen that the possibility of $Y(4660)$ taken as the state
$\psi(4^3S_1)$ or $\psi(6^3S_1)$ can't be excluded completely.
Fixing the masses at $M=4643$ MeV, we obtain
\begin{eqnarray}
\Gamma[\psi(4^3S_1)\rightarrow\Lambda_c\bar{\Lambda}_c]\sim6.57~\text{MeV},
\end{eqnarray}
and
\begin{eqnarray}
\Gamma[\psi(6^3S_1)\rightarrow\Lambda_c\bar{\Lambda}_c]\sim0.84~\text{MeV}.
\end{eqnarray}
The predicted decay widths are large enough to be observed in
experiments. Moreover,
\begin{eqnarray}
\frac{\Gamma[\psi(4^3S_1)\rightarrow\Lambda_c\bar{\Lambda}_c]}
{\Gamma[\psi(5^3S_1)\rightarrow\Lambda_c\bar{\Lambda}_c]}\sim2.7.
\end{eqnarray}
The branching ratio of $\psi(4^3S_1)$ decaying into
$\Lambda_c\bar{\Lambda}_c$ pair is larger than that of
$\psi(5^3S_1)$.

Meanwhile, if the $Y(4660)$ resonance corresponds to $\psi(6^3S_1)$, the branching ratio is predicted to be
\begin{eqnarray}
\mathcal{B}[\psi(6^3S_1)\rightarrow\Lambda_c\bar{\Lambda}_c]\sim1\%.
\end{eqnarray}
This branching ratio is the smallest, while it is quite large
compared to the ratio
($\mathcal{O}(10^{-3})\sim\mathcal{O}(10^{-5})$) of other charmonium
states decaying into the baryon-antibaryon
pair~\cite{Patrignani:2016xqp}.

In addition, we also plot the decay width of the $\psi(4^3S_1)$ and
$\psi(6^3S_1)$ as a function of the mass in the range of
$M=(4580-4800)$ MeV in Fig.~\ref{SWAVE}. The variation curves
between the partial decay width and the mass for these two states
are similar to that for $\psi(5^3S_1)$.

\begin{figure*}[htpb]
\centering \epsfxsize=15.5 cm \epsfbox{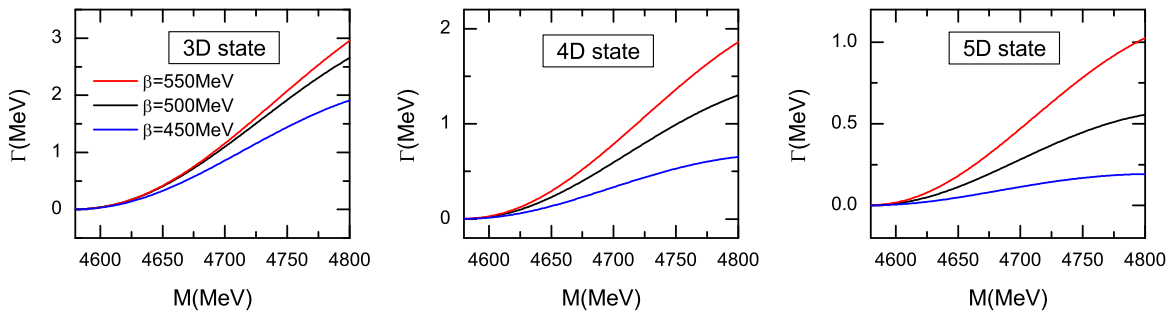} \caption{The
variation of the $\Lambda_c\overline{\Lambda}_c$ decay width with
the mass of the D-wave vector charmonium. The blue, black, and red
lines correspond to the predictions with different values of the
harmonic oscillator strength $\beta=450$, 500, and 550 MeV,
respectively.}\label{DWAVE}
\end{figure*}

In brief, we have calculated the $\Lambda_c\bar{\Lambda}_c$ partial
decay widths of the three $S$-wave states $\psi(4^3S_1)$,
$\psi(5^3S_1)$, and $\psi(6^3S_1)$ with the QPC model. According to
our predictions, the $\Lambda_c\bar{\Lambda}_c$ decay width can
reach up to a few MeV. If $Y(4660)$ is a vector charmonium, it is
very likely to be found in the $\Lambda_c\bar{\Lambda}_c$ channel.

\subsection{$D$ wave}
The predicted mass of the state $\psi(4^3D_1)$ is listed in
Table~\ref{MASS}. This state is a good candidate of the $Y(4660)$
resonance. So it is necessary to investigate the decay properties of
$\psi(4^3D_1)$.

In the same way, we fix the mass of $\psi(4^3D_1)$ at $M=4643$ MeV
firstly. Then, we obtain the partial decay width
\begin{eqnarray}
\Gamma[\psi(4^3D_1)\rightarrow\Lambda_c\bar{\Lambda}_c]\sim0.19~\text{MeV}.
\end{eqnarray}
This width seems not large, but it is enough to be
observed in this decay channel in experiments. Moreover,
the branching ratio is predicted to be
\begin{eqnarray}
\mathcal{B}[\psi(4^3D_1)\rightarrow\Lambda_c\bar{\Lambda}_c]\sim0.3\%.
\end{eqnarray}

We also plot the variation of the $\Lambda_c\bar{\Lambda}_c$ decay
width as a function of the mass in Fig.~\ref{DWAVE}. From the
figure, the partial decay width for $\psi(4^3D_1)$ decaying into the
$\Lambda_c\bar{\Lambda}_c$ pair is less than $\sim1.3$ MeV in the
range of $(4580-4800)$ MeV.

Furthermore, we analyze the decay properties of the states
$\psi(3^3D_1)$ and $\psi(5^3D_1)$, and collect their predicted
masses in Table~\ref{MASS}. From the table, the theoretical masses
are either about ($100\sim200$) MeV lighter or heavier than the mass
of $Y(4660)$ in PDG~\cite{Patrignani:2016xqp}. We also study the
decay properties of the two states in this work.

Taking the masses of the $\psi(3^3D_1)$ and $\psi(5^3D_1)$ as
$M=4643$ MeV, we get that the partial decay widths are
\begin{eqnarray}
\Gamma[\psi(3^3D_1)\rightarrow\Lambda_c\bar{\Lambda}_c]\sim0.33~\text{MeV}
\end{eqnarray}
and
\begin{eqnarray}
\Gamma[\psi(5^3D_1)\rightarrow\Lambda_c\bar{\Lambda}_c]\sim0.09~\text{MeV}.
\end{eqnarray}
If the $Y(4660)$ were the state $\psi(5^3D_1)$, one would be very
hard to observe $Y(4660)$ in the $\Lambda_c\bar{\Lambda}_c$ channel.

The predicted masses of the states $\psi(3^3D_1)$ and $\psi(5^3D_1)$
certainly have a large uncertainty, which may bring uncertainties to
our theoretical results. To investigate this effect, we plot the
partial decay widths of these two states as functions of the masses
in Fig.~\ref{DWAVE} as well.

The $\Lambda_c\bar{\Lambda}_c$ decay widths of the $D$-wave states
are less than one MeV. The $\Lambda_c\bar{\Lambda}_c$ decay width
ratio between the $S$-wave states and $D$-wave states is
$\mathcal{O}(10)$. If $Y(4660)$ turns out to be a $S$-wave
state, it has a good potential to be observed in the
$\Lambda_c\bar{\Lambda}_c$ channel.

\subsection{The effect of $\beta$}

We have considered six excited vector charmonium states around 4.6
Gev and investigated their $\Lambda_c\bar{\Lambda}_c$ partial decay
width. In the present work, all of the theoretical predictions are
obtained with the parameter $\beta=500$ MeV. However, the harmonic
oscillator parameter $\beta$ for the excitation between the charm
quarks in initial charmonium system is not determined precisely,
which bares a large uncertainty. To investigate the uncertainties of
the parameter $\beta$, we further consider the decay properties as a
function of the mass with two different $\beta$ values: $\beta=450$,
550 MeV. The numerical results are shown in
Figs.~\ref{SWAVE}-\ref{DWAVE}. One notes that the bigger $\beta$
value leads to a larger decay width. Our main predictions hold in a
reasonable range of the parameter $\beta$.

\section{Summary}\label{suma}

In the present work, we calculate the $\Lambda_c\bar{\Lambda}_c$
partial decay width of the excited vector charmonium around 4.6 GeV,
including $\psi(4S,~5S,~6S)$ and $\psi(3D,~4D,~5D)$. The
$\Lambda_c\bar{\Lambda}_c$ mode is not kinematically allowed for the
charmonium states below 4.6 GeV. This OZI allowed mode provides a
new tool to explore the higher charmonium, which will be produced
abundantly at Belle-II. We extend the original $^3P_0$ model and
consider the creation of two light $q\bar q$ pairs from the vacuum,
which is the first attempt along this direction in literatures up to
our knowledge.

Based on our calculations, the decay widths of the $S$-wave states
decaying into the $\Lambda_c\bar{\Lambda}_c$ pair are about a few
MeV, while the $\Lambda_c\bar{\Lambda}_c$ decay widths of the
$D$-wave states are less than one MeV. The
$\Lambda_c\bar{\Lambda}_c$ decay width ratio between the $S$-wave
states and the $D$-wave states is $\mathcal{O}(10)$. If the
$Y(4660)$ is one of the $S$-wave states considered in this work, it
may be observed in the $\Lambda_c\bar{\Lambda}_c$ channel. Moreover,
if the enhancement $Y(4630)$ reported by Belle collaboration in
$\Lambda_c\bar{\Lambda}_c$ invariant-mass distribution is the same
structure as the $Y(4660)$, the $Y(4660)$ is very likely to be a
$S$-wave charmonium state. On the other hand, it will be very
difficult to observe the excited D-wave vector charmonium in the
$\Lambda_c\bar{\Lambda}_c$ channel. In other words, the
$\Lambda_c\bar{\Lambda}_c$ mode can be used to pin down the internal
structure of the vector charmonium.

\section*{Acknowledgements }

We would like to thank Xiao-Lin Chen and Wei-Zhen Deng for very
helpful discussions. This work is supported by the National Natural
Science Foundation of China under Grants No. 11621131001,
11575008 and 11775078. This work is also in part supported by China Postdoctoral
Science Foundation under Grant No.~2017M620492 and by National Key
Basic Research Program of China (2015CB856700).

\begin{appendix}
\section{The Amplitude calculations}
The harmonic oscillator wave functions for the ground charmed
baryons $B(C)$ in our calculation are
\begin{eqnarray}
\psi^{B(C)}(0,~0,~0,~0)&=&3^{\frac{3}{4}}\Bigg
(\frac{1}{\pi\alpha_{\rho}^2}\Bigg)^{\frac{3}{4}}\Bigg
(\frac{1}{\pi\alpha_{\lambda}^2}\Bigg)^{\frac{3}{4}} \nonumber\\
&&\times\exp\Bigg[-\frac{P_{\rho}^2}{2\alpha_{\rho}^2}-\frac{P_{\lambda}^2}{2\alpha_{\lambda}^2}\Bigg],
\end{eqnarray}
where
$\mathbf{p}^{B(C)}_{\rho}=\frac{1}{\sqrt{2}}(\mathbf{p}_{3(2)}-\mathbf{p}_{5(4)})$
and
$\mathbf{p}^{B(C)}_{\lambda}=\frac{1}{\sqrt{6}}(\mathbf{p}_{3(2)}+\mathbf{p}_{5(4)}-2\mathbf{p}_{1(2)})$.

The ground state wave function of the meson is
\begin{eqnarray}
\Psi(0,~0)=\Bigg(\frac{1}{\pi\beta^2}\Bigg)^{\frac{3}{4}}
\exp\Bigg[-\frac{(\mathbf{p}_1-\mathbf{p}_2)^2}{8\beta^2}\Bigg]
\end{eqnarray}

Since all the final states are in the $S$-wave states in this
calculations, the momentum space integration
$I^{M_{L_A},m,m'}_{M_{L_B}M_{L_C}}(\mathbf{p})$ can be further
expressed as $\prod(M_{L_A},m,m')$.

For the $1S$ charmonium state decay:
\begin{eqnarray}
\Pi(0,~0,~0)&=&\Bigg
(\frac{1}{\pi\alpha_{\rho}^2}\Bigg)^{\frac{3}{2}}\Bigg
(\frac{1}{\pi\alpha_{\lambda}^2}\Bigg)^{\frac{3}{2}}\Bigg
(\frac{1}{\pi\beta^2}\Bigg)^{\frac{3}{4}}\exp\Bigg[-(\lambda_4-\frac{\lambda^2_3}{4\lambda_2})P^2_B\Bigg] \nonumber\\
&&\times\frac{\pi^2}{16}\big(\frac{1}{\lambda_1\lambda_2}\big)^{\frac{3}{2}}\big(12\varpi^2P_B^2-\frac{3}{\lambda_1}+\frac{1}{\lambda_2}\big),\\
\Pi(0,1,-1)&=&\Bigg
(\frac{1}{\pi\alpha_{\rho}^2}\Bigg)^{\frac{3}{2}}\Bigg
(\frac{1}{\pi\alpha_{\lambda}^2}\Bigg)^{\frac{3}{2}}\Bigg
(\frac{1}{\pi\beta^2}\Bigg)^{\frac{3}{4}}\exp\Bigg[-(\lambda_4-\frac{\lambda^2_3}{4\lambda_2})P^2_B\Bigg] \nonumber\\
&&\times\frac{\pi^2}{16}\big(\frac{1}{\lambda_1\lambda_2}\big)^{\frac{3}{2}}\big(\frac{3}{\lambda_1}-\frac{1}{\lambda_2}\big)\nonumber\\
&=&\Pi(0,-1,1).
\end{eqnarray}

For the $1D$ charmonium state decay:
\begin{eqnarray}
\Pi(0,~0,~0)&=&\Bigg
(\frac{1}{\pi\alpha_{\rho}^2}\Bigg)^{\frac{3}{2}}\Bigg
(\frac{1}{\pi\alpha_{\lambda}^2}\Bigg)^{\frac{3}{2}}\Bigg
(\frac{1}{\beta^2}\Bigg)^{\frac{7}{4}}\exp\Bigg[-(\lambda_4-\frac{\lambda^2_3}{4\lambda_2})P^2_B\Bigg] \nonumber\\
&&\times\pi^{\frac{5}{4}}\big(\frac{1}{\lambda_1\lambda_2}\big)^{\frac{3}{2}}\Bigg(\frac{\sqrt{3}}{2}\zeta^2\varpi^2P^4_B
+\frac{\zeta^2P^2_B}{8\sqrt{3}\lambda_2}\nonumber\\
&& -\frac{\sqrt{3}\zeta^2P^2_B}{8\lambda_1}-\frac{\zeta\varpi P^2_B}{\sqrt{3}\lambda_2}+\frac{\sqrt{3}}{36\lambda^2_2}\Bigg),\\
\Pi(0,~1,-1)&=&\Bigg
(\frac{1}{\pi\alpha_{\rho}^2}\Bigg)^{\frac{3}{2}}\Bigg
(\frac{1}{\pi\alpha_{\lambda}^2}\Bigg)^{\frac{3}{2}}\Bigg
(\frac{1}{\beta^2}\Bigg)^{\frac{7}{4}}\exp\Bigg[-(\lambda_4-\frac{\lambda^2_3}{4\lambda_2})P^2_B\Bigg] \nonumber\\
&&\times\pi^{\frac{5}{4}}\big(\frac{1}{\lambda_1\lambda_2}\big)^{\frac{3}{2}}\Bigg(
\frac{\sqrt{3}\zeta^2P^2_B}{8\lambda_1}-\frac{\zeta^2P^2_B}{8\sqrt{3}\lambda_2}+\frac{\sqrt{3}}{72\lambda^2_2}\Bigg)\nonumber\\
&=&\Pi(0,-1,1),\\
\Pi(1,-1,0)&=&\Bigg
(\frac{1}{\pi\alpha_{\rho}^2}\Bigg)^{\frac{3}{2}}\Bigg
(\frac{1}{\pi\alpha_{\lambda}^2}\Bigg)^{\frac{3}{2}}\Bigg
(\frac{1}{\beta^2}\Bigg)^{\frac{7}{4}}\exp\Bigg[-(\lambda_4-\frac{\lambda^2_3}{4\lambda_2})P^2_B\Bigg] \nonumber\\
&&\times\pi^{\frac{5}{4}}\big(\frac{1}{\lambda_1\lambda_2}\big)^{\frac{3}{2}}\Bigg(
\frac{\zeta \varpi P^2_B}{4\lambda_2}-\frac{1}{24\lambda^2_2}\Bigg)\nonumber\\
&=&\Pi(1,0,-1),\\
\Pi(2,-1,-1)&=&\Bigg
(\frac{1}{\pi\alpha_{\rho}^2}\Bigg)^{\frac{3}{2}}\Bigg
(\frac{1}{\pi\alpha_{\lambda}^2}\Bigg)^{\frac{3}{2}}\Bigg
(\frac{1}{\beta^2}\Bigg)^{\frac{7}{4}}\exp\Bigg[-(\lambda_4-\frac{\lambda^2_3}{4\lambda_2})P^2_B\Bigg] \nonumber\\
&&\times\pi^{\frac{5}{4}}\big(\frac{1}{\lambda_1\lambda_2}\big)^{\frac{3}{2}}
\frac{\sqrt{2}}{24\lambda^2_2}.\\
\end{eqnarray}
Here,
\begin{eqnarray}
\lambda_1&=&\frac{1}{\alpha^2_{\rho}},\\
\lambda_2&=&\frac{1}{\alpha^2_{\lambda}}+\frac{1}{3\beta^2},\\
\lambda_3&=&-\frac{\sqrt{6}}{9\beta^2},\\
\lambda_4&=&\frac{1}{18\beta^2},\\
   \varpi&=&\frac{1}{3}-\frac{\lambda_3}{2\sqrt{6}\lambda_2},\\
    \zeta&=&\frac{1}{3}+\frac{\lambda_3}{\sqrt{6}\lambda_2},
\end{eqnarray}
for the above expressions. $|P_B|$ reads as
\begin{eqnarray}
|P_B|=\frac{\sqrt{\big(m^2_A-(m^2_B+m^2_C)\big)\big(m^2_A-(m^2_B-m^2_C)\big)}}{2m_A}.
\end{eqnarray}

With the amplitudes for the $1S$ states decaying into two $S$-wave
final states, we can obtain the radially and orbitally excited
states' amplitudes which are related to the lowest radial or orbital
states by differentiation~\cite{Liu:2011yp},
\begin{eqnarray}
\mathcal{M}_{4S}&=&\frac{1}{3\sqrt{35}}\big(15\beta\frac{\partial}{\partial\beta}+6\beta^2\frac{\partial^2}{\partial\beta^2}
+2\beta^3\frac{\partial^3}{\partial\beta^3}\big)\mathcal{M}_{1S},\\
\mathcal{M}_{5S}&=&\frac{1}{18\sqrt{70}}\big(63+72\beta\frac{\partial}{\partial\beta}+96\beta^2\frac{\partial^2}{\partial\beta^2}
+24\beta^3\frac{\partial^3}{\partial\beta^3}\nonumber\\
&&+4\beta^4\frac{\partial^4}{\partial\beta^4}\big)\mathcal{M}_{1S},\\
\mathcal{M}_{6S}&=&\frac{1}{45\sqrt{77}}\big(\frac{675}{2}\beta\frac{\partial}{\partial\beta}+240\beta^2\frac{\partial^2}{\partial\beta^2}
+120\beta^3\frac{\partial^3}{\partial\beta^3}\nonumber\\
&&+20\beta^4\frac{\partial^4}{\partial\beta^4}+2\beta^5\frac{\partial^5}{\partial\beta^5}\big)\mathcal{M}_{1S},\\
\mathcal{M}_{3D}&=&\frac{1}{3\sqrt{14}}\big(7+2\beta\frac{\partial}{\partial\beta}+2\beta^2\frac{\partial^2}{\partial\beta^2}
\big)\mathcal{M}_{1D},\\
\mathcal{M}_{4D}&=&\frac{1}{3\sqrt{231}}\big(27\beta\frac{\partial}{\partial\beta}+6\beta^2\frac{\partial^2}{\partial\beta^2}
+2\beta^3\frac{\partial^3}{\partial\beta^3}\big)\mathcal{M}_{1D},\\
\mathcal{M}_{5D}&=&\frac{1}{6\sqrt{6006}}\big(231+120\beta\frac{\partial}{\partial\beta}+144\beta^2\frac{\partial^2}{\partial\beta^2}
+24\beta^3\frac{\partial^3}{\partial\beta^3}\nonumber\\
&&+4\beta^4\frac{\partial^4}{\partial\beta^4}\big)\mathcal{M}_{1D}.
\end{eqnarray}

\end{appendix}

\end{document}